\documentclass[preprint2]{aastex}
\newcommand{\hband}{$H_{160}$}
\newcommand{\jband}{$J_{125}$}
\newcommand{\zband}{$z_{850}$}
\newcommand{\iband}{$i_{814}$}

\slugcomment{Accepted for Publication in ApJ}

\shorttitle{When did round disk galaxies form?}
\shortauthors{Takeuchi et al.}

\begin{document}


\title{When did round disk galaxies form?}


\author{T. M. Takeuchi\altaffilmark{}, and K. Ohta\altaffilmark{}}
\affil{Department of Astronomy, Kyoto University,
    Kyoto 606-8502, Japan}
\email{ohta@kusastro.kyoto-u.ac.jp}

\author{S. Yuma\altaffilmark{}}
\affil{Institute for Cosmic Ray Research, The University of Tokyo, Kashiwa 277-8582, Japan}
%\email{aastex-help@aas.org}

\and

\author{K. Yabe\altaffilmark{}}
\affil{National Astronomical Observatory of Japan, Mitaka 181- 8588, Japan}


%\altaffiltext{1}{Visiting Astronomer, Cerro Tololo Inter-American Observatory.
%CTIO is operated by AURA, Inc.\ under contract to the National Science
%\altaffiltext{2}{Society of Fellows, Harvard University.}
%\altaffiltext{3}{present address: Center for Astrophysics,
%    60 Garden Street, Cambridge, MA 02138}
%\altaffiltext{4}{Visiting Programmer, Space Telescope Science Institute}
%\altaffiltext{5}{Patron, Alonso's Bar and Grill}


\begin{abstract}
When and how galaxy morphology such as disk and bulge
seen in the present-day universe emerged is still not clear.
In the universe at $z\gtrsim 2$, galaxies with various morphology 
are seen, and star-forming galaxies at $z\sim2$ show
an  intrinsic shape of bar-like structure.
Then, when did round disk structure form?
Here we take a simple and straightforward approach to see the epoch 
when a round disk galaxy population emerged by constraining 
the intrinsic shape statistically based on  
apparent axial ratio distribution of galaxies.
We derived the distributions of the apparent axial ratios in the
rest-frame optical light ($\sim 5000$ \AA)  of 
star-forming main sequence galaxies at $2.5>z>1.4$,
$1.4>z>0.85$, and $0.85>z>0.5$, and 
found that the apparent axial ratios of them show  peaky distributions
at $z\gtrsim0.85$, while a rather flat distribution at the lower redshift.
By using a tri-axial model ($A>B>C$) for the  intrinsic shape,
we found the best-fit models give the peaks of the $B/A$ distribution
 of $0.81\pm0.04$, $0.84\pm0.04$, and $0.92\pm0.05$  at $2.5>z>1.4$,
$1.4>z>0.85$, and $0.85>z>0.5$, respectively.
The last value is close to the local value of 0.95.
Thickness ($C/A$) is $\sim0.25$ at all the redshifts 
and is close to the local value (0.21).
The results indicate the shape of the star-forming galaxies
in the main sequence changes gradually, and the round disk
is established at around $z\sim0.9$.
Establishment of the round disk may be due to a cease of violent 
interaction of galaxies or a growth of a bulge and/or
 a super-massive black hole resides at the center of a 
galaxy which dissolves the bar structure. 
\end{abstract}





\section{Introduction}

When and how  did disk galaxies form?　 
At $z \lesssim 1$,  disk galaxies that can be classified with
the Hubble's tuning fork exist \citep[e.g.,][]
{sch95,abr96,lil98,sca07,sar07}; 
they are  identified with their apparent morphology, 
and/or their surface brightness distribution, and/or 
empirical indices describing such as central condensation 
and asymmetry. 
Meanwhile, in the universe at $z\gtrsim3$, 
galaxies with various morphology are seen,  
and many of them show the presence of clumps and/or
irregularity, though the morphology is traced in the
rest-frame UV  \citep[e.g.,][]{gia96,ste96}.
Hence, the emergence of morphology classified with the Hubble's tuning
fork is considered to be around at $z\sim 1-3$, and formation of disk 
structure would also be expected at the epoch. 

The Wide Field Camera 3 (WFC3) on the {\it Hubble Space Telescope} ($HST$)
enables us to study morphology of galaxies in the rest-frame optical
light  well up to at $z\sim 3$.
\citet{con11} studied morphology evolution based on eye inspection.
They found the disk population is increasing with redshift gradually
and the population is very rare at $z>2$.
By using apparent morphology, similar trend is reported 
by \citet{cam11} for galaxies with a stellar mass
range of $10^{10-12} M_{\odot}$ and by \citet{tal14}.
\citet{mor13} also studied the morphology evolution 
with eye inspection and
a clearer emergence of the disk population at $z\sim2$ is seen
after correcting for the statistical morphology misclassification.
 
However, from the optical surface brightness distribution, 
the fraction of the disk galaxies is higher at $z>2$ 
\citep{mor13,bru12,bui13}.
If the S\'{e}rsic index ($n$) less than 2.5 is
regarded to be an indicator of disk galaxies,  
the  fraction of the disk population
is larger at $z>2$ and is decreasing with redshift,
though the details of the evolutional trend is not so simple;
e.g., if irregular galaxies are included in the sample, the disk
fraction is more constant against redshift \citep{bui13},
or the fraction is rather constant for less massive galaxies
($< 10^{10.5} M_{\odot}$) \citep{mor13}.
\citet{bru12} tried to model the two-dimensional 
surface brightness distribution with two components of
S\'{e}rsic index of $n=1$ and $n=4$, and found a fraction of
the disk ($n=1$ component) dominated galaxies decreases at
$z\sim2$ rather sharply for stellar mass larger than $10^{10.5}
M_{\odot}$.

Thus the opposite views are obtained in terms of
the disk evolution.
%, though the S\'{e}rsic index may not be a good indicator for disk galaxies.  
Other parameters characterizing the morphological features 
such as central condensation, asymmetry, Gini, M20, etc 
seem not to be suitable to trace the morphology of galaxies 
in distant universe \citep{wan12,mor13,tal14}.

On the other hand, studies of internal kinematics in star-forming
galaxies such as BM/BX and star-forming BzK (sBzK) galaxies at
$z\sim2$  revealed about one third of them show a clear rotation 
kinematics \citep{for09}.
This suggests we witness disk galaxies and they may be progenitor
of present-day disk galaxies, though a fraction of them
may evolve into the present-day elliptical galaxies through
major merges,
particularly those with large stellar mass.
Another piece of the suggestion  for the disk progenitor scenario 
stems from the clustering property of sBzK galaxies;
sBzK galaxies with $K < 23$ mag are considered to be
a progenitor of present-day disk galaxies with regard to 
their weak clustering strength \citep{hay07}.
It would be  reasonable to suppose most of the sBzK galaxies 
in this magnitude range will evolve into the present-day disk galaxies,
though the brighter sBzK galaxies may be a progenitor of elliptical galaxies. 

Motivated by the finding of the presence of many rotating  disk 
among sBzK galaxies and clustering nature of them,  
we took a simple and straightforward  approach to see
whether the intrinsic shape of faint sBzK galaxies at $z\sim2$ is 
disk-like or not statistically based on 
apparent axial ratio distribution. 
We studied the surface brightness distribution and the apparent 
axial ratio in both rest-frame UV and optical light for sBzK galaxies 
with $K_{\rm S} <24$ mag  \citep{yum11,yum12}.
\citet{yum11} and \citet{yum12} revealed 
the surface brightness distribution of them is characterized by 
S\'{e}rsic index of $n\sim 1$, and the half-light radius and
surface stellar mass density is similar to those in the
present-day disk galaxies \citep{bar05}.
However, it was turned out that the intrinsic shape of the sBzK galaxies
is not a round disk  (round means intrinsic axial ratio of 
$\sim 0.9-1.0$); 
their intrinsic shape is bar-like or oval  (intrinsic axial ratio
$\sim 0.6-0.8)$
with the thickness  similar to that of the present-day disk galaxies.
The intrinsic bar-like structure of star-forming galaxies
at $z\sim2$ is also pointed out by \citet{law12}.
It should be noted here that this bar-like structure does not
imply a direct progenitor of the local barred galaxies as discuss later.

Then, when did round disk galaxies appear after $z\sim2$?
Emergence of the round disk population should reflect the 
evolution process of the disk structure.
Hence, revealing evolution of the intrinsic shape of the galaxies
is expected to give us an insight to  physical process of the galaxy evolution.
In this paper, we study the intrinsic shape of main sequence
galaxies with $K_{\rm S}<24$ mag in the redshift 
range from 2.5 to 0.5
using the same method as Yuma et al. (2011) and Yuma et al. (2012)
to see the redshift that the round disk galaxy population emerged. 
In the subsequent section, data sources and sample galaxies
are summarized.  
In \S 3, we derive the surface brightness distributions and
axial ratios of the galaxies.
Then, in \S 4, we examine the distributions of the apparent
axial ratios and constrain the intrinsic shape of the galaxies.
Conclusion and discussion are given in \S 5.   
We adopt a cosmology with $H_0 = 70$ km s$^{-1}$ Mpc$^{-1}$,
$\Omega_{\rm m} = 0.3$, and $\Omega_{\Lambda} = 0.7$.
All magnitudes in this paper are given in the AB magnitude system.

\section{Data sources and sample galaxies}\label{data}

Sample galaxies in this study are collected from 
the Great Observatories Origins Deep Survey-South (GOODS-S)\footnote{
http://www.sci.edu/science/goods}  \citep{dic03}
and the Subaru XMM-Newton Deep Survey (SXDS)  \citep{fur08}, 
which almost overlaps Ultra Deep Survey (UDS) in 
UK Infrared Telescope Infrared Deep Sky Survey (UKIDSS)\citep{law07}.  
The data sources are the same as those used by \citet{yum12}
and the details are described in the paper. 
Here we briefly summarize them.  
In GOODS-S region, near infrared (NIR) images ($J$-, $H$-, and 
$K_{\rm S}$-band images) were taken from GOODS/ISAAC (Infrared
Spectrometer And Array Camera) data release final version \citep{ret10}. 
Five sigma limiting $K_{\rm S}$-band magnitude is 24.4 mag.  
NIR images taken with WFC3 on $HST$ with F125W and F160W
filters were obtained  from CANDELS \citep[Cosmic Assembly Near-infrared Deep
Extragalactic Legacy Survey;][]{gro11,koe11}\footnote{
http://candels.ucolick.org/index.html} data release v0.5. 
Optical images  were taken from  ver. 2.0 data products of GOODS 
$HST$/ACS(Advanced Camera for Surveys)  treasure program
 ($B_{435}$, $V_{660}$, $i_{775}$, and $z_{850}$) \citep{gia04} 
and from ESO/GOODS program ($U$ and $R$) \citep{non09}. 
Mid-infrared (MIR) images at 3.6 $\mu$m and 4.5 $\mu$m taken 
with Infrared Array Camera (IRAC) on the {\it Spitzer Space Telescope} ($SST$) 
were from data release 1 and 2 obtained in Spitzer Legacy Science
program. 
 In SXDS/UDS region, NIR images ($J$, $H$, and $K_{\rm S}$) 
were taken from data release ver. 8 of UKIDSS UDS.  
Five sigma limiting magnitude is $K_{\rm S} = 24.6$ mag. 
NIR images with $HST$/WFC3  with  F125W and F160W filters were  
obtained from CANDELS data release v1.0. 
Optical images were taken from SXDS project ($B$, $V$, $R_{\rm C}$,
$i^{\prime}$, and $z^{\prime}$)  \citep{fur08}.  
We used $HST$/ACS images taken with F814W filter (\iband).
MIR 3.6$\mu$m and 4.5$\mu$m images were taken from 
Spitzer Public Legacy Survey of UKIDSS UDS (SpUDS; PI: J. Dunlop).
We also use the far-UV and near-UV data taken from Galaxy Evolution Explorer
\citep[$GALEX$;][]{mar05}
% (GALEX; Martin et al. 2005, ApJ, 619, L1) 
archived image (GR6) and 
the $u$-band data of the Canada-France-Hawaii Telescope Legacy Survey (CFHTLS) 
wide field as supplemental information, though the limiting magnitudes
are shallow.

Object detection was made based on $K_{\rm S}$-band image by using
SE{\scriptsize XTRACTOR} \citep{ber96}.
 We made photometry using these data 
% to IRAC 4.5$\mu$m-bands
for each source detected in $K_{\rm S}$ images 
at the same position and derived total magnitudes 
using SE{\scriptsize XTRACTOR}.  
For the images taken with ACS  and IRAC, we applied 
a correction  for the aperture photometry 
to obtain a total magnitude following the manner by Yuma et al. (2012).  
Spectral energy distribution (SED) was constructed for each object 
from the photometry in all the bands. 
Then photometric redshifts were derived by using {\scriptsize HYPERZ} \citep{bol00}. 
 Comparisons with spectroscopic redshifts available showed  $\Delta
 z/(1+z)$  of $\sim0.05$ and $\sim0.03$ in GOODS-S and SXDS, respectively.  
With the photometric redshifts, stellar mass, star-formation rate
 (SFR),  and color excess ($E(B-V)$) were derived for each object 
through SED fitting  by using  {\it SEDfit} program \citep{saw12}. 
The {\it SEDfit} employs population synthesis code of BC03\citep{bru03}.  
Salpeter initial mass function with a mass range of $0.1-100$
$M_{\odot}$ and the solar metallicity were assumed.  
Extinction law used was that by Calzetti et al. (2000). 
More details are described by Yuma et al. (2011, 2012) 
and Yabe et al. (2012, 2014).

Using these data, we selected galaxies brighter than $K_{\rm S} = 24.0$ mag. 
We divided the galaxies in three epochs of  $2.5>z>1.4$,
$1.4>z>0.85$, and  $0.85>z>0.5$  based on the photometric redshifts;
 these epochs are chosen to take 
a comparable duration ($\sim 2$ Gyr) of the epochs in the cosmic age and 
to keep a reasonable sample size.  

\begin{figure*}[!htb]
\epsscale{1.70}
\plotone{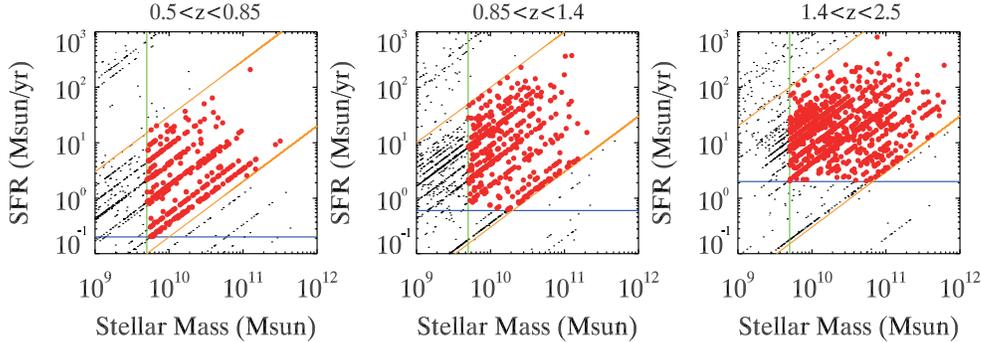}
\caption{
Star formation rate versus stellar mass of the sample galaxies at three
epochs:  
From right to left,  $2.5>z>1.4$, $1.4>z>0.85$, and $0.85>z>0.5$.  
Large red points are sample in this study  and solid lines in
each panel show the selection criteria} (see text).  
\label{fig1}
\end{figure*}

 Figure \ref{fig1} shows distribution of the SFR against 
stellar mass in each epoch. 
 As like in other works \citep[e.g.,][]{noe07,dad07}, the SFRs are larger in 
more massive galaxies 
and they make a sequence  referred as main sequence.  
The main sequence shows a gradual cosmological evolution down to $z\sim0$, 
where we see disk galaxies in the sequence.  
Most of our sample galaxies at $2.5>z>1.4$ are also sBzK galaxies.  
%It is worth noting here non-star-forming galaxies are located 
%in far below of the main sequence and they are considered to be 
%progenitor of  present-day elliptical galaxies. 
Galaxies in the most massive part of the main sequence at the higher redshifts
may soon stop star formation and evolve into passive elliptical
galaxies according to a simple theoretical expectation
\citep[e.g.,][]{bou10}.
But most of the sample galaxies occupy less
massive part of the main sequence  and they are expected to 
reside in the main sequence at the present epoch.
A  branch seen above the main sequence shows very high SFR 
at a fixed stellar mass, and they are considered to be 
violently forming stars often under strong galaxy interaction 
\citep[e.g.,][]{dad07}.  
Thus we selected star-forming galaxies in the main sequence bracketed
 by thresholds shown in Figure \ref{fig1}.
The stellar mass limit at $5 \times 10^{9} M_{\odot}$ is to avoid the
inclusion of Magellanic-type galaxies that would tend to  show bar-like
structure even in the local universe.
A star-forming galaxy with a stellar mass of $\sim 5 \times 10^{9}
M_{\odot}$  at $z=2.5$ is a progenitor of the Milky Way class galaxy
(stellar mass of $5 \times 10^{10} M_{\odot}$), if we adopt 
the abundance matching in terms of the number density \citep{van13}.
Ranges of the specific SFR (sSFR) we use are 
$3 \times 10^{-11} - 2 \times 10^{-8}$ yr$^{-1}$ ($z=1.4-2.5$),
$3 \times 10^{-11} - 1 \times 10^{-8}$ yr$^{-1}$ ($z=0.85-1.4$),
and
$2 \times 10^{-11} - 3 \times 10^{-9}$ yr$^{-1}$ ($z=0.5-0.85$).
SFR limit is just  an additional threshold; 
SFR $> 2.0 M_{\odot}$ yr$^{-1}$ ($z=1.4-2.5$),
$> 0.6 M_{\odot}$ yr$^{-1}$    ($z=0.85-1.4$),
and
$> 0.2 M_{\odot}$ yr$^{-1}$ ($z=0.5-0.85$).
%Many of them apparently show a disk-like structure in the rest-frame
% optical band, but 
A fraction (typically $10 - 15$\%) of them show 
multiple-structure in the rest-frame optical  $HST$ images;
if an object detected by SE{\scriptsize XTRACTOR} locates
within a radius of median size (FWHM) of the galaxies
in each epoch, we regard them as multiple.
Such galaxies are not adequate to examine the shape and 
we discarded them in this study. 

\section{Two-dimensional surface brightness fitting}

We made two-dimensional surface brightness distribution 
fitting to the galaxies selected above.
The images were taken with $HST$ in the rest-frame optical 
wavelength $(\sim0.5\mu$m):
 F160W(\hband) for $2.5>z>1.4$, 
 F125W($J_{125}$) for $1.4>z>0.85$, 
and 
 F850LP(\zband) in GOODS-S and F814W(\iband) in SXDS for $0.85>z>0.5$.  
Magnitude ranges for the sample galaxies in three epochs 
were  virtually  \hband $<$ 24 mag and \jband $<$ 24 mag,
 and  \zband $<$ 25 mag in GOODS-S and 
\iband $<$ 24 mag in SXDS.
Pixel scales after drizzle were  $0.^{\prime\prime}06$,
 $0.^{\prime\prime}06$, and $0.^{\prime\prime}03$  
in \hband, \jband, and \zband/\iband, respectively.

The radial distribution is fitted with S\'{e}rsic profile: 
$ I(r)= I_e$ exp$[-\kappa_n \{(r/r_e)^{1/n} -1\}]$, 
where  $r_e$ is the half light radius, 
$I_e$ is the surface brightness at $r_e$, and $n$ characterizes 
the shape of the profile (S\'{e}rsic index).  
%S\'{e}rsic index of 1 corresponds to an exponential distribution and 
%is typical for disk galaxies in the local universe,
% while $n$ of 4 is typical for present-day elliptical galaxies. 
 Two dimensional surface brightness fitting was made by using
 GALFIT version 3 \citep{pen10}.  
Initial guess parameters on the central position,  axial ratio, 
position angle,  and $r_e$ are taken from the output parameters
of SE{\scriptsize XTRACTOR}.  
Initial S\'{e}rsic index was set to be 1.5, which does not 
affect the final results significantly.  
Point spread function in each image was constructed from $5\sim20$
 isolated unsaturated stars.  

\begin{figure}[!htb]
%\plottwo{f2.eps}{f2_color.eps}
\epsscale{0.9}
\plotone{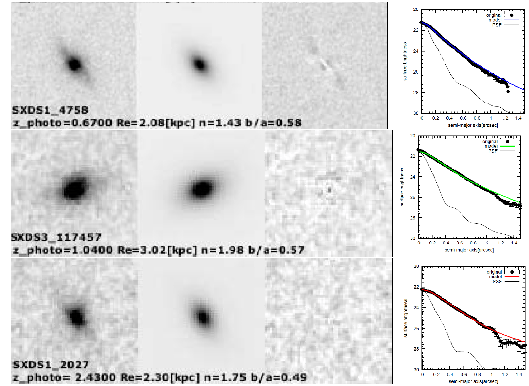}
\caption{
Example of the surface brightness distribution fitting 
at  $0.85>z>0.5$(upper), $1.4>z>0.85$(middle), 
and $2.5>z>1.4$ (lower).  
The images are taken  with $HST$ \iband (upper), WFC3 \jband (middle), 
and WFC3 \hband (lower).  
In each row,  from left to right, original image, 
best-fit model image, residual, and radial surface brightness profiles 
along the major axis are shown. 
In the rightmost panel, the observed surface brightness and 
the best-fitted profile smoothed with point spread function 
is shown with  circles with error and solid curve, respectively. 
 Point spread function is  shown with thin solid curve.
}\label{fig2}
\end{figure}

Figure \ref{fig2} shows example of the profile fitting.  
As seen in the figure, the surface brightness profiles extend to
$\sim1$ arcsec in radius corresponding to $\sim8$ kpc at $z=1\sim2$ 
and $\sim6$ kpc at $z=0.5$. 
The surface brightness we use to fit the model reaches  
$23.5 \sim 25$ mag arcsec$^{-2}$ in the rest-frame $V$ band 
after correcting for the cosmological dimming.
This level of the surface brightness can be regarded as
 the outskirt of galaxies, and is deep enough to confront with
the shape of local disk galaxies.  
Considering the fading of galaxy luminosity to the present epoch, 
the value would be $\sim 1$ mag arcsec$^{-2}$ fainter at $z \sim 0$ 
\citep[e.g.,][]{bri98,mil11}. 
We obtained $r_e$, $n$, and apparent axial ratio ($b/a$, 
where $a$ is a major axis and $b$ is a minor axis)  
for each galaxy.  
About 60-70\%  of the sample galaxies show $n=0.5-2.5$. 
We then examined the distribution of $r_e$ and stellar mass of 
the sample galaxies with $n=0.5-2.5$.  
In each epoch, the distribution of $r_e$ against stellar mass is 
similar to that for local disk galaxies \citep{bar05}. 

\begin{figure}[!htb]
%\plottwo{f2.eps}{f2_color.eps}
\includegraphics[angle=0,scale=0.35]{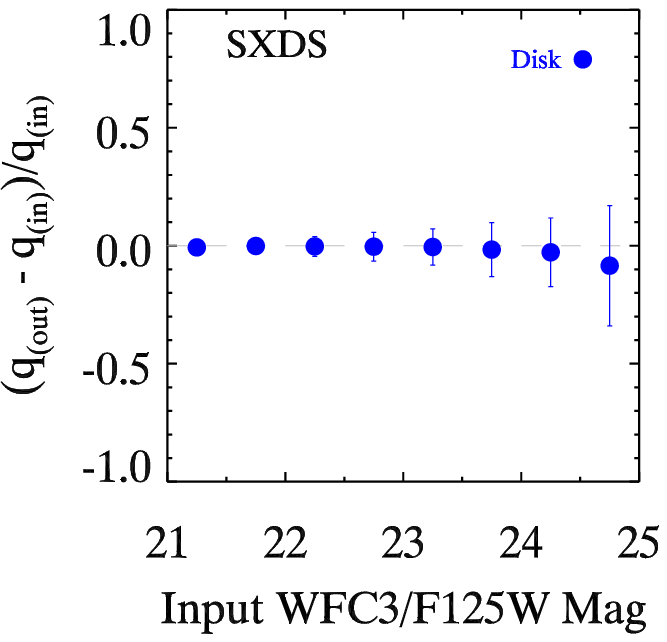}
\includegraphics[angle=0,scale=0.35]{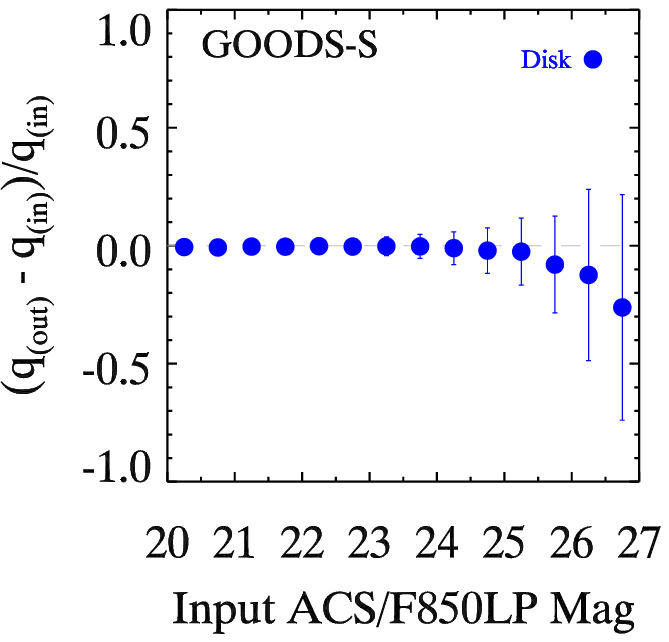}
\includegraphics[angle=0,scale=0.35]{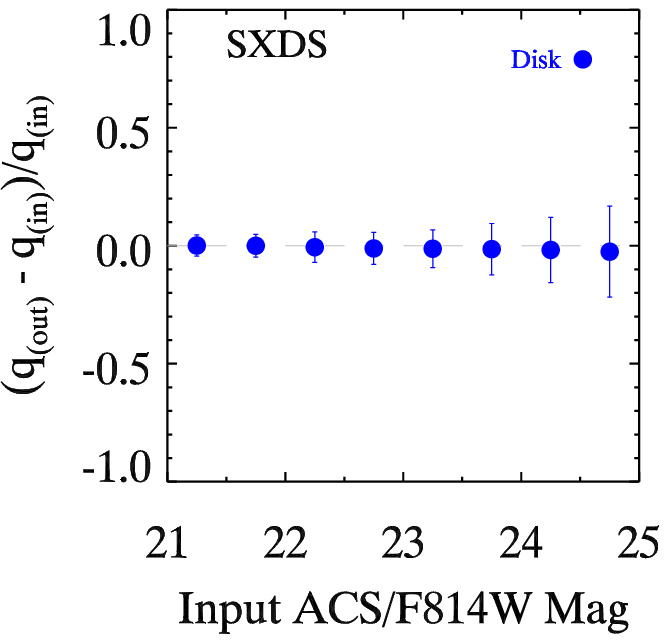}
%\epsscale{0.8}
%\plotone{fig3.eps}
\caption{
Accuracy of the obtained axial ratio against magnitude of 
input artificial object in \jband (left),  \zband (middle), and \iband
(right).
Error bars represent 1$\sigma$ dispersion of the distribution. 
}\label{fig3}
\end{figure}

Accuracy for the S\'{e}rsic index  as well as for the axial
ratio was examined by putting artificial objects into the images 
used for the analysis. 
The artificial objects were created by using ARTDATA/MKOBJECTS in
IRAF. 
Mock galaxies with  S\'{e}rsic  index of 1 or 4 were 
generated in both GOODS-S and SXDS with half-light radius of 
$0.^{\prime\prime}1-1.^{\prime\prime}0$ in a magnitude range of 
$20-25$ mag.
  The ranges of input parameters are chosen to mimic  those 
of the sample galaxies. 
 Axial ratio and position angle are randomly chosen and the galaxies 
are put into \hband,  \jband, \zband, and \iband\, images randomly.  
As we did for the real objects, we first used SE{\scriptsize XTRACTOR}
to derive position, position angle, axial ratio, and FWHM. 
We then performed GALFIT with these parameters as well as $n=1.5$
 as the initial guess. 
Results for  \hband (GOODS-S, SXDS/UDS) were presented by 
Yuma et al. (2012).   
Accuracy for the S\'{e}rsic index in \hband\, image 
is good enough ($<10$ \%) in this magnitude range in case of $n=1$. 
Even if the intrinsic $n$ is 4, the misclassification into $n=1$ disk 
is negligible, though the accuracy is slightly worse in \zband\, image
\citep{yum11,yum12}.
We show the accuracy of the axial ratio derived
for \jband (SXDS/UDS), \zband (GOODS-S), and \iband
(SXDS/UDS)  in case of $n=1$ in Figure \ref{fig3}. 
In all the panels,  the obtained axial ratios agree very well with 
those of input value in the bright magnitude range, 
and the accuracy is getting worse for the fainter objects.  
However, at the faintest magnitude level of the sample galaxies
in each band, 
the accuracy of the obtained axial ratio 
is less than  5\% on average, which  is smaller than the bin size
of the histograms of  the axial ratio distribution  described below. 
The accuracy of axial ratio in \hband (GOODS-S, SXDS/UDS) is
 presented by \citet{yum12}, and it is
comparable to those in  other bands.
Detection rate in $K_{\rm S}$-band image against the apparent axial ratio 
was also examined  
and the rate is  almost constant 
%($\sim 90$ \%)  
in the range of $b/a=0.2$ and 1.0
 \citep{yum11}.
\vspace{0.4cm}

\section{Axial ratio distribution}

\subsection{Distribution of apparent axial ratios} 
In order to see whether they have round disk structure, 
we take a simple and straightforward approach; 
i.e., we constrain the intrinsic shape of these galaxies 
statistically based on distribution of  apparent axial ratios 
of the sample galaxies at three epochs.  
Figure \ref{fig4} (upper row) shows the apparent axial ratio distributions
 of the sample galaxies with  $n=0.5-2.5$ among galaxies 
selected in \S\ref{data}.  
If the intrinsic shape of disk galaxies is an infinitesimally 
thin circular disk, the apparent axial ratio distribution is 
exactly flat.  
In fact, the distribution for  $\sim300,000$ local disk galaxies
 is almost flat except for  around at $b/a \sim 0$ and $\sim 1$
\citep{pad08}, 
since real disks have a finite thickness and are not perfectly round.   
As seen in the right panel ($2.5>z>1.4$) of the upper row of 
Figure \ref{fig4}, 
the axial ratio distribution peaks at $\sim 0.45$ and sharply 
decreases to both sides.  
In the middle panel ($1.4>z>0.85$), the distribution is slightly 
flatter, but still shows a slight peaky distribution.  
The histograms suggest  these galaxies do not have a round 
disk structure.  
Meanwhile in the left panel ($0.85>z>0.5$), the distribution
 is rather flat and is close to that for the local disk galaxies,
 suggesting they have the intrinsic shape of round disk.

%\vspace{1.4cm}

\subsection{Intrinsic axial ratio distribution}
 To constrain  the intrinsic shape quantitatively, 
we employ a tri-axial model with axial lengths of  $A>B>C$ 
following the method by Ryden (2004).  The face-on ellipticity 
($\epsilon  = 1 -  B/A$) is assumed to be described by log-normal 
distribution with mean of $\mu (=$ ln $\epsilon$) 
and dispersion of $\sigma$, while the edge-on thickness 
($C/A$) is assumed to be described with Gaussian distribution 
with mean of $\mu_\gamma$ and dispersion of $\sigma_\gamma$.  
With this parameter set, the expected apparent axial ratio 
distribution seen from random viewing angles can be calculated
and can be compared with the observed apparent axial ratio distribution.  

The model axial ratio distributions were calculated in a range of 
$\mu = -3.95 \sim -0.05$,  $\sigma = 0.2 \sim 2.0$,
$\mu_{\gamma}=0.1\sim0.98$, and  $\sigma_{\gamma} = 0.01 \sim 0.35$,  
with a step of $\Delta \mu =0.15,　\Delta \sigma=0.15,　
\Delta \mu_{\gamma} =0.02$,  and $\Delta \sigma_{\gamma}=0.02$.  
With these parameter sets, we searched for the best-fit parameter set 
using $\chi^2$ method with $1\sigma$ statistics by Gehrels
(1986). 
To estimate the uncertainty of the best-fit parameters, 
we made Monte Carlo realization of the observed histogram 
of the axial ratio 1000  times  
and derived the best-fit parameter set for each realization.  
We took 68\% confidence level as the uncertainty of the model
parameter. 
The peak value of the distribution of  intrinsic $B/A$ is 
given by $1-$exp($\mu - \sigma^2$).  
The uncertainty on the peak $B/A$ was calculated based on  
the propagation of errors with the uncertainty of $\mu$ and $\sigma$
　mentioned above.
 
\begin{figure*}[!htb]
\epsscale{1.70}
\plotone{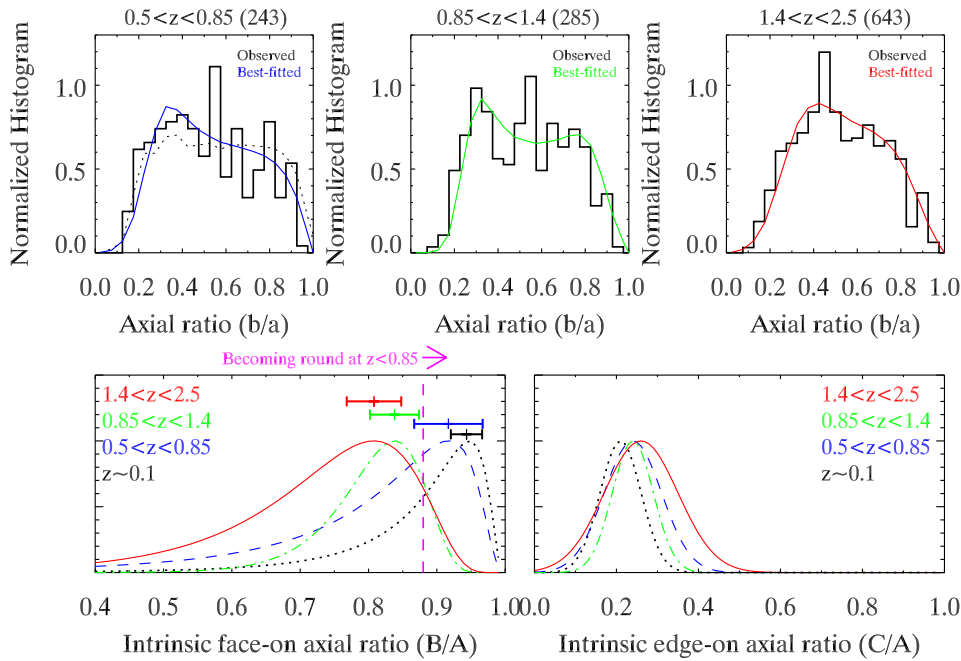}
\caption{
 Normalized histograms of the observed axial ratios at each epoch (upper
panels):From right to left,  $2.5>z>1.4$, $1.4>z>0.85$, and $0.85>z>0.5$.  
Numbers in vertical axis are multiplied by 10.
Number in the parenthesis at the top of each panel refers to the
sample size.   
Best-fit model distributions are shown with solid curves. 
Axial distribution of local disk galaxies from the Sloan Digital Sky 
Survey Data Release 6 by Padilla \& Strauss (2008) is also shown with
dotted curve (left panel).  
Distribution of best-fit intrinsic axial ratios of $B/A$ and $C/A$
(bottom panels). 
Red solid, green dot dashed, and blue dashed curves refer to the intrinsic model axial 
distribution at   $2.5>z>1.4$, $1.4>z>0.85$, and
$0.85>z>0.5$, respectively. 
Black dotted curves show the distribution for local disk galaxies 
\citep{pad08}.
Y-axis is an arbitrary unit.  
Peaks for the $B/A$ distribution are shown by tick marks 
with uncertainty above the curves.
\label{fig4}}
\end{figure*}

The best-fit models are shown with solid curves 
in Figure \ref{fig4} (upper row), and the best-fitted parameters
as well as the uncertainties are shown in Table 1.  
Resulting distributions of intrinsic axial ratios at the epochs are 
shown in Figure \ref{fig4} (bottom row). 
The best-fit model $B/A$ peaks at  $0.81\pm0.04,  0.84\pm0.04$, 
and $0.92\pm0.05$ at  $2.5>z>1.4, 1.4>z>0.85$, and $0.85>z>0.5$, 
respectively, and the last value of $B/A$ is very close to that 
for the local disk galaxies (0.94) \citep{pad08}.  
The mean $C/A$ is $\sim 0.25$ at all the epochs and is close to 
that of the local disk galaxies (0.21) \citep{pad08}.  
The intrinsic shape of the star-forming main sequence galaxies 
at $2.5>z>0.85$ is bar-like or oval, 
while they have round disk structure at $z<0.85$, 
i.e., the emergence of the round disk population is $z \sim 0.9$, 
though the precise redshift determination is difficult due to the 
the gradual cosmological evolution and small sample size.  
%We note here that the selection of redshift bins and of the 
%criteria for the sample galaxies (SFR, specific SFR, $n$)  
%are not sensitive to the results except for the stellar mass cut;  
%including much smaller mass galaxies changes the axial ratio
%distribution.  

\begin{table*}[!htb]
\begin{center}
\caption{Best-fit parameter sets for the tri-axial model.\label{tbl-1}}
\begin{tabular}{ccccc}
\tableline\tableline
    & Local\tablenotemark{*} & $z=0.5-0.85$ & $z=0.85-1.4$ & $z=1.4-2.5$ \\
\tableline
$\mu$ & $-2.33^{+0.13}_{-0.13}$ & $-1.85^{+0.35}_{-0.18}$ &
$-1.70^{+0.12}_{-2.30}$ & $-1.40^{+0.08}_{-0.08}$ \\
$\sigma$ & $0.79^{+0.16}_{-0.16}$ & $0.80^{+0.19}_{-0.60}$ &
   $0.35^{+0.19}_{-0.13}$ & $0.50^{+0.17}_{-0.18}$ \\
$\mu_{\gamma}$ &
$0.21^{+0.02}_{-0.02}$ & $0.24^{+0.02}_{-0.02}$ & 
 $0.24^{+0.01}_{-0.01}$ & $0.26^{+0.02}_{-0.16}$ \\
$\sigma_{\gamma}$ & $0.050^{+0.015}_{-0.015}$ &
$0.07^{+0.16}_{-0.06}$ & $0.05^{+0.01}_{-0.01}$ &  $0.09^{+0.01}_{-0.08}$ \\
\tableline
\end{tabular}
%% Any table notes must follow the \end{tabular} command.
\tablenotetext{*}{Padilla \& Strauss (2008)}
\end{center}
\label{tbl}
\end{table*}

%%% precise error version %%%
%\begin{table*}
%\begin{center}
%\caption{Best-fit parameter sets for the tri-axial model.\label{tbl-1}}
%\begin{tabular}{ccccc}
%\tableline\tableline
%    & Local\tablenotemark{*} & $z=0.5-0.85$ & $z=0.85-1.4$ & $z=1.4-2.5$ \\
%\tableline
%$\mu$ & $-2.33^{+0.13}_{-0.13}$ & $-1.85^{+0.348}_{-0.177}$ &
%$-1.70^{+0.1176}_{-2.30}$ & $-1.40^{+0.075}_{-0.075}$ \\
%$\sigma$ & $0.79^{+0.16}_{-0.16}$ & $0.80^{+0.185}_{-0.600}$ &
%   $0.35^{+0.1901}_{-0.1309}$ & $0.50^{+0.1653}_{-0.1824}$ \\
%$\mu_{\gamma}$ &
%$0.21^{+0.02}_{-0.02}$ & $0.24^{+0.022}_{-0.015}$ & 
% $0.24^{+0.01}_{-0.01}$ & $0.26^{+0.0167}_{-0.16}$ \\
%$\sigma_{\gamma}$ & $0.050^{+0.015}_{-0.015}$ &
%$0.07^{+0.156}_{-0.060}$ & $0.05^{+0.01}_{-0.01}$ &  $0.09^{+0.0141}_{-0.08}$ \\
%\tableline
%\end{tabular}
%% Any table notes must follow the \end{tabular} command.
%\tablenotetext{*}{Padilla \& Strauss (2008)}
%\end{center}\label{tbl1}
%\end{table*}

\subsection{Robustness of the results}
In order to examine the robustness of the results, 
we constructed samples by changing selection criteria slightly 
with respect to $K_{\rm S}$-band magnitude, stellar mass, sSFR, 
\iband/\zband/\jband/\hband\,  magnitude, and S\'{e}rsic index, 
and made the same fitting to these samples.

If we take samples in each epoch with 0.5 mag brighter or fainter 
in $K_{\rm S}$ band, virtually all the best-fit parameters are 
within 1$\sigma$  error.  
The peak values of the intrinsic axial distribution ($B/A$) agree 
with the best-fit value within 1 $\sigma$ error.　　
If we take a massive sample (stellar mass larger than $1.0 \times
10^{10} M_{\odot}$),   the results mostly agree within 1 $\sigma$ error.  
If we include less massive galaxies (stellar mass larger than 
$3 \times 10^9 M_{\odot}$), peak values of $B/A$ agree within
the error, 
but the resulting parameters for the highest-redshift bin
change more than 1 $\sigma$.  
It should be noted that even in local universe such less massive 
galaxies often show bar-like structure as like Magellanic irregular 
galaxies.  
We also test the dependence on sSFR  by taking 
$3 \times 10^{-11}$ yr$^{-1} <$ sSFR $< 2 \times 10^{-8}$  yr$^{-1}$
 at $z=2.5-1.4$,  $3 \times 10^{-11}$ yr$^{-1} <$ sSFR $< 5 \times
 10^{-9}$  yr$^{-1}$ at $z=1.4-0.85$,  and $3 \times 10^{-11}$
 yr$^{-1} <$ sSFR $<  7 \times 10^{-9}$  yr$^{-1}$ at $z=0.85-0.5$.  
Almost all the obtained best-fit parameters as well as the 
peak values of $B/A$ agree within 1$\sigma$ error.
If we take the threshold SFR 2-3 times larger, the resulting
parameters are almost within 1 $\sigma$.
  The axial ratios are derived by using \iband, \zband, \jband, and 
\hband\,   images.  
More accurate axial ratios are expected to be obtained with the 
brighter sample.  
Hence we take about 0.5 mag brighter samples (\iband $<$ 23.5 mag, 
\zband $<$ 24.5 mag,  \jband $<$ 23.5 mag, and \hband $<$ 23.5 mag)  
from the  $K_{\rm S}<24.0$ mag sample.  
The resulting peak  axial ratios are within or marginally within 1
$\sigma$ uncertainty,  and the conclusion in this study does not
change.
 We also make samples with S\'{e}rsic index of $0.5-1.5$ and
 $0.5-3.5$,  and find again the conclusion does not change. 

Dust obscuration may affect the axial ratio distribution;
the redder disk galaxies tend to show the smaller  axial ratios
in the local universe as well as at $z =0.6-0.9$
 \citep[e.g.,][]{pad08,pat12}.
This trend is considered to be due to the larger 
dust obscuration in more inclined galaxies.
In order to see the effect, we divided the sample into
two subsamples with smaller or larger color excess ($E(B-V)$)
at each epoch.
The color excess was derived from the SED fitting mentioned in
\S 2, and we divided the sample at around the median $E(B-V)$
of 0.2 or 0.25 mag.
Resulting histograms of the axial ratios show that 
a slight excess at smaller axial ratio ($b/a \sim 0.2-0.4$)
is seen at $z= 0.5 \sim 0.85$,
but such a clear trend is not seen at the higher redshifts.
We made the model fitting to the subsamples.
About half of the best fit $\mu$ and $\sigma$ values agree with
those of the total sample within the uncertainty for the total sample.
However, it should be noted that the uncertainty for the subsamples
is large due to the smaller sample size.
We also derived the peak $B/A$ with the subsamples.
The peak $B/A$ values in the smaller color excess (larger color
excess)
are 0.81 (0.93) at $2.5>z>1.4$, 0.91 (0.78) at $1.4>z>0.85$,
and 0.99 (0.97) at $0.85>z>0.5$.
Thus the increasing trend of $B/A$ with decreasing $z$ is still seen,
except for the $B/A$ for the larger color excess at $2.5>z>1.4$.

\section{Conclusion and discussion}
We constrain the intrinsic shape of the star-forming 
main sequence galaxies at $z=2.5$ to 0.5 
based on the distributions of
the apparent axial ratios  by employing the
tri-axial model of $A>B>C$.
At $z \gtrsim 0.9$, the intrinsic shape of them is
bar-like or oval with $B/A \sim 0.8$, while
at $z \lesssim 0.9$, $B/A$ is 0.92 that is close to the
value for the local galaxies (0.95).
Hence, the emergence of the round disk can be regarded 
to be  $z\sim0.9$,
though the precise redshift
determination is difficult due to the gradual evolution and
the  small sample size.
The thickness ($C/A$) is about 0.25 and does not show
significant cosmological evolution down to $z\sim0$.
Similar result has been obtained very recently by
adopting a slightly different model for the intrinsic shape;
the triaxiality of star-forming galaxies with stellar mass
larger than $10^{9.5} M_{\odot}$ is large
at $z \gtrsim 1$, but small at $z \lesssim 1$ \citep{van14}. 
It is worth noting the bar-like structure we see at the high 
redshifts is unlikely to be a direct progenitor of bar structure
 in the present-day barred galaxies, because the apparent axial 
ratios are obtained mainly from outside beyond the bar region 
in the present-day barred galaxies; the bar region  in
a present-day barred galaxy is brighter than
$\sim 22$ mag arcsec$^{-2}$ in $V$ band  \citep{oht90,oht07}. 
Furthermore, large barred galaxies are disappearing as redshift increases
\citep{van96}: 
Its fraction among disk galaxies is only 10-20\% at $z = 0.8 \sim 1.0$
\citep{she08,mel14}. 

At $z\gtrsim3$, most of star-forming galaxies show S\'{e}rsic index 
of $\sim1$ in the rest-frame UV \citep{rav06}
and in the rest-frame optical  \citep{aki08}.
However, the distribution of the apparent axial ratios
is not flat, suggesting the elongated intrinsic structure
\citep{rav06}.
The results obtained here reveal  that although the disk-like structure 
is seen among star-forming galaxies at $z\gtrsim1$, its intrinsic shape is
 bar-like or oval.
Physical mechanism to make such structure is not clear, 
but they may be bar-like due to galaxy interaction and/or minor merge
\citep{nog87,ber04}. 
Or the baryonic matter might have been  already highly concentrated 
in a dark matter halo, resulting into the bar instability \citep{ost73}.  

Then how did they become round?  
Numerical simulations show  a central mass condensation 
(either of a bulge and/or a super-massive black hole (SMBH)) 
with a mass fraction of 1-10\% of a disk dissolves the  
bar-like structure \citep{ath05,hoz12}. 
Thus the emergence of round disk population may be related to 
the growth of a central bulge and/or a SMBH.  
According to the simulations  by \citet{ath05}, 
the bar-like structure gradually evolves 
into a round disk with a time scale of a few Gyr,
which is comparable to the elapsed time from  $z=1.5$ to 0.9 
($\sim 2$ Gyr).
However, eye inspection of our sample galaxies at $z=0.85-0.5$
suggests that bulge dominated galaxies are not the majority.
The two-component model fit to discriminate bulge and
disk in a galaxy is not straightforward  and  it is beyond our scope in
this study. 
AGN number density peaks at $z=1-2$ \citep[e.g.,][]{ued03,has05},
 suggesting the rapid growth of SMBHs during the epoch. 
The growth of a SMBH may also play a role for  dissolving 
the bar structure, although the mass fraction may not be
large enough.  
Alternatively, the gradual cease of galaxy interaction may make
 a round disk.  
If this is the case, the epoch of the round disk population
at $z\sim0.9$ would indicate the major galaxy interaction 
mostly ceased around at this redshift and the galaxy morphology
is rather stable after this.  
Further studies of  link between  intrinsic shape of disk and  
the growth of a bulge/SMBH as well as galaxy interaction are 
desirable to disclose the evolution of disk galaxies.

\acknowledgments
We thank the referee for the useful comments.
KO's efforts are supported by the Grant-in-Aid for Scientific Research 
 (24540230) and 
the Grant-in-Aid for Scientific Research on Innovative Area (24103003) 
from the Japan Society for the Promotion of Science.
This work is partially based on observations taken by the
CANDELS Multi-Cycle Treasury Program with the NASA/ESA $HST$, which is
operated by the Association of Universities for Research in Astronomy,
Inc., under NASA contract NAS5-26555. Parts of the observations in
GOODS-S have been carried out using the Very Large Telescope at the
ESO Paranal Observatory under Program ID 168.A-0485.
Based on observations obtained with MegaPrime/MegaCam, 
a joint project of CFHT and CEA/IRFU, 
at the Canada-France-Hawaii Telescope (CFHT) which is operated 
by the National Research Council (NRC) of Canada,
the Institut National des Science de l'Univers of the 
Centre National de la Recherche Scientifique (CNRS) of France,
 and the University of Hawaii.
 This work is based in part on data products produced at Terapix
 available 
at the Canadian Astronomy Data Centre 
as part of the Canada-France-Hawaii Telescope Legacy Survey, 
a collaborative project of NRC and CNRS.

\clearpage


\begin{thebibliography}{}
\bibitem[Abraham et al.(1996)]{abr96} Abraham, R. G., Tanvir, N. R.,
Santiago, B. X., et al.(1996), \mnras, 279, L47
\bibitem[Akiyama et al.(2008)]{aki08} Akiyama, M., Minowa, 
Y., Kobayashi, N., et al.\ 2008, \apjs, 175, 1
\bibitem[Athanassoula et al.(2005)]{ath05} Athanassoula, E., Lambert, J. C., \& Dehnen, W.  2005, \mnras, 363, 496
\bibitem[Barden et al.(2005)]{bar05}  Barden, M., Rix, H., Somerville, R. S., et al. 2005, \apj, 635, 959
\bibitem[Berentzen et al.(2004)]{ber04}  Berentzen, I., Athanassoula,
E., Heller, C. H. \& Fricke, K.J. 2004, \mnras, 347,  220
\bibitem[Bertin \& Arnouts(1996)]{ber96} Bertin, E., \& Arnouts, S. 1996, \aaps, 117, 393
\bibitem[Bolzonella et al.(2000)]{bol00} Bolzonella, M., Miralles,
J.-M. \& Pell\'{o}, R. 2000, \aap, 363, 476
\bibitem[Bouch{\'e} et al.(2010)]{bou10} Bouch{\'e}, N., Dekel, A., Genzel, R., et al.\ 2010, \apj, 718, 1001 
\bibitem[Brinchmann et al.(1998)]{bri98} Brinchmann, J., Abraham, R., Schade, D., et al. 1998, \apj, 499, 112
\bibitem[Bruce et al.(2012)]{bru12} Bruce, V. A., Dunlop, J. S., Cirasuolo, M.,  et al. 2012, \mnras, 427, 1666
\bibitem[Bruzual \& Charlot(2003)]{bru03} Bruzual, G., \& Charlot, S.\ 2003, \mnras, 344, 1000 
\bibitem[Buitrago et al.(2013)]{bui13} Buitrago, F., Trujillo, I., Conselice, C. J., \& H\"{a}u\ss ler, B. 2013, \mnras, 428, 1460
\bibitem[Calzetti et al.(2000)]{cal00} Calzetti, D., Armus, L., Bohlin, R. C., et al. 2000, \apj, 533,  682
\bibitem[Cameron et al.(2011)]{cam11} Cameron, E., Carollo, C. M., Oesch, P. A., et al. 2011, \apj, 743, 146
\bibitem[Conselice et al.(2011)]{con11} Conselice, C. J., Bluck, A. F. L., Ravindranath, S., et al. 2011, \mnras, 417, 2270
\bibitem[Daddi et al.(2007)]{dad07} Daddi, E., Dickinson, M., Morrison,
G., et al. 2007, \apj, 670, 156
\bibitem[Dickinson et al.(2003)]{dic03} Dickinson, M., 
Giavalisco, M., \& GOODS Team 2003, in The Mass of Galaxies at Low and
High Redshift, ed. R. Bender \& A. Renzini (Berlin:Springer), 324 
\bibitem[F\"{o}rster Schreiber et al.(2009)]{for09} F\"{o}rster Schreiber,
N. M., Genzel, R., Bouch\'{e}, N., et al.  2009, \apj, 706, 1364
\bibitem[Furusawa et al.(2008)]{fur08} Furusawa, H., Kosugi, G., Akiyama, M., et al. 2008, \apjs, 176, 1
\bibitem[Gehrels(1986)]{geh86} Gehrels, N.  1986, \apj, 303, 336
\bibitem[Giavalisco et al.(1996)]{gia96} Giavalisco, M., Steidel, C. C., \& Macchetto, F. D.  1996,  \apj, 470, 189
\bibitem[Giavalisco et al.(2004)]{gia04} Giavalisco, M., Ferguson, H. C., Koekemoer, A. M., et  al. 2004, \apjl, 600,  L93
\bibitem[Grogin et al.(2011)]{gro11} Grogin, N. A., Kocevski, D. D., Faber, S. M., et al. 2011, \apjs, 197, 35
\bibitem[Hasinger et al.(2005)]{has05} Hasinger, G., Miyaji, T., \& Schmidt, M. 2005, \aap, 441, 417
\bibitem[Hayashi et al.(2007)]{hay07} Hayashi, M., Shimasaku, K., Motohara, K., et al. 2007, \apj, 660, 72
\bibitem[Hozumi(2012)]{hoz12} Hozumi, S. 2012, \pasj,  64, 5
\bibitem[Koekemoer et al.(2011)]{koe11} Koekemoer, A. M., Faber, S. M., Ferguson, H. C., et al. 2011, \apjs, 197,  36
\bibitem[Law et al.(2012)]{law12} Law, D. R., Steidel, C. C., Shapley, A. E., et al. 2012, \apj, 745,  85
\bibitem[Lawrence et al.(2007)]{law07} Lawrence, A., Warren, S. J., Almaini, O., et al. 2007, \mnras,  379,  1599
\bibitem[Lilly et al.(1998)]{lil98} Lilly, S., Schade, D., Ellis, R., et
al.  1998, \apj, 500, 75
\bibitem[Martin et al.(2005)]{mar05} Martin, D.~C., Fanson, 
J., Schiminovich, D., et al.\ 2005, \apjl, 619, L1 
\bibitem[Melvin et al.(2014)]{mel14} Melvin, T., Masters, K., 
Lintott, C., et al.\ 2014, \mnras, 438, 2882 
\bibitem[Miller et al.(2011)]{mil11} Miller, S. H., Bundy, K., Sullivan, M., Ellis, R. S., \& Treu, T. 2011, \apj, 741, 115
\bibitem[Mortlock et al.(2013)]{mor13} Mortlock, A., Conselice, C. J., Hartley, W. G., et al. 2013, \mnras, 433, 1185
\bibitem[Noeske et al.(2007)]{noe07} Noeske, K. G., Weiner, B. J., Faber, S. M., et al. 2007, \apjl, 660,  L43
\bibitem[Noguchi(1987)]{nog87} Noguchi, M. 1987, \mnras, 228, 635
\bibitem[Nonino et al.(2009)]{non09} Nonino, M., Dickinson, M., Rosati,
P., et al. 2009, \apjs, 183, 244
\bibitem[Ohta et al.(1990)]{oht90} Ohta, K., Hamabe, M., \& Wakamatsu, K. 1990, \apj,  357,  71
\bibitem[Ohta et al.(2007)]{oht07} Ohta, K., Aoki, K., Kawaguchi, T.,
\& Kiuchi, G.  2007, \apjs, 169,  1
\bibitem[Ostriker \& Peebles(1973)]{ost73}  Ostriker, J. P., \&
Peebles, P. J. E. 1973, \apj, 186, 467
\bibitem[Padilla and Strauss(2008)]{pad08}  Padilla, N. D., \& Strauss,
M. A. 2008, \mnras, 388, 1321
\bibitem[Patel et al.(2012)]{pat12} Patel, S.~G., Holden, 
B.~P., Kelson, D.~D., et al.\ 2012, \apjl, 748, L27 
\bibitem[Peng et al.(2010)]{pen10} Peng, C. Y., Ho, L. C.,  Impey, C. D. \& Rix, H.-W.  2010, \aj, 139, 2097
\bibitem[Ravindranath et al.(2006)]{rav06} Ravindranath, S., 
Giavalisco, M., Ferguson, H.~C., et al.\ 2006, \apj, 652, 963 
\bibitem[Retzlaff et al.(2010)]{ret10} Retzlaff, J., Rosati, P., Dickinson, M., et al. 2010, \aap, 511, 50
\bibitem[Ryden(2004)]{ryd04} Ryden, B. S. 2004, \apj, 601, 214
\bibitem[Sargent et al.(2007)]{sar07} Sargent, M. T., Carollo, C. M.,
Lilly, S., et al.  2007, \apjs, 172, 434
\bibitem[Sawicki(2012)]{saw12}  Sawicki, M. 2012, \pasp, 124, 1208
\bibitem[Scarlata et al.(2007)]{sca07} Scarlata, C., Carollo, C. M.,
Lilly, S., et al.  2007, \apjs, 172, 406
\bibitem[Schade et al.(1995)]{sch95} Schade, D., Lilly, S. J.,
Crampton, D., et al.  1996, \apjl, 451, L1
\bibitem[Sheth et al.(2008)]{she08} Sheth, K., Elmegreen, D.,
Elmegreen, B. G., et al. 2008, \apj, 675, 1141
\bibitem[Steidel et al.(1996)]{ste96} Steidel, C.~C., 
Giavalisco, M., Dickinson, M., \& Adelberger, K.~L.\ 1996, \aj, 112, 352 
\bibitem[Talia et al.(2014)]{tal14} Talia, M., Cimatti, A., Mignoli, M., et al. 2014, \aap, 562, A113
\bibitem[Ueda et al.(2003)]{ued03}  Ueda, Y.,  Akiyama, M.,  Ohta, K., \& Miyaji, T. 2003, \apj, 598,  886
\bibitem[van den Bergh et al.(1996)]{van96}  van den Bergh, S.,
Abraham, R. G., Ellis, R. S., et al. 1996, \aj, 112, 359
\bibitem[van der Wel et al.(2014)]{van14} van der Wel, A., 
Chang, Y.-Y., Bell, E.~F., et al.\ 2014, \apjl, 792, L6 
\bibitem[van Dokkum et al.(2013)]{van13} van Dokkum, P.~G., Leja, J., Nelson, E.~J., et al.\ 2013, \apjl, 771, L35 
\bibitem[Wang et al.(2012)]{wan12} Wang, T., Huang, J.-S., Faber, S. M., et al. 2012, \apj, 752, 134
\bibitem[Yabe et al.(2012)]{yab12} Yabe, K., Ohta, K., Iwamuro, F.,  et al. 2012, \pasj, 64, 60
\bibitem[Yabe et al.(2014)]{yab14} Yabe, K., Ohta, K., Iwamuro, F.,   et al. 2014, \mnras, 437, 3647
\bibitem[Yuma et al.(2011)]{yum11} Yuma, S., Ohta, K., Yabe, K., Kajisawa, M., \& Ichikawa, T.  2011, \apj, 736, 92
\bibitem[Yuma et al.(2012)]{yum12} Yuma, S., Ohta, K., \& Yabe, K. 2012, \apj,  761, 19


%\bibitem[Auri\`ere(1982)]{aur82} Auri\`ere, M.  1982, \aap,    109, 301
%\bibitem[Canizares et al.(1978)]{can78} Canizares, C. R.,
%    Grindlay, J. E., Hiltner, W. A., Liller, W., \&
%    McClintock, J. E.  1978, \apj, 224, 39
%\bibitem[Djorgovski \& King(1984)]{djo84} Djorgovski, S.,
%    \& King, I. R.  1984, \apjl, 277, L49
%\bibitem[Hagiwara \& Zeppenfeld(1986)]{hag86} Hagiwara, K., \&
%    Zeppenfeld, D.  1986, Nucl.Phys., 274, 1
%\bibitem[Harris \& van den Bergh(1984)]{har84} Harris, W. E.,
%    \& van den Bergh, S.  1984, \aj, 89, 1816
%\bibitem[H\`enon(1961)]{hen61} H\'enon, M.  1961, Ann.d'Ap., 24, 369
%\bibitem[Heiles \& Troland(2003)]{heiles03} Heiles, C. \& Troland, T. H., 2003,% \apjs, preprint doi:10.1086/381753
%\bibitem[Kim, Ostricker, \& Stone(2003)]{kim03} Kim, W.-T.,  Ostriker, E., \& S%tone, J. M., 2003, \apj, 599, 1157
%\bibitem[King(1966)]{kin66}  King, I. R.  1966, \aj, 71, 276
%\bibitem[King(1975)]{kin75}  King, I. R.  1975, Dynamics of
%    Stellar Systems, A. Hayli, Dordrecht: Reidel, 1975, 99
%\bibitem[King et al.(1968)]{kin68}  King, I. R., Hedemann, E.,
%    Hodge, S. M., \& White, R. E.  1968, \aj, 73, 456
%\bibitem[Kron et al.(1984)]{kro84} Kron, G. E., Hewitt, A. V.,
%    \& Wasserman, L. H.  1984, \pasp, 96, 198
%\bibitem[Lynden-Bell \& Wood(1968)]{lyn68} Lynden-Bell, D.,
%    \& Wood, R.  1968, \mnras, 138, 495
%\bibitem[Newell \& O'Neil(1978)]{new78} Newell, E. B.,
%    \& O'Neil, E. J.  1978, \apjs, 37, 27
%\bibitem[Ortolani et al.(1985)]{ort85} Ortolani, S., Rosino, L.,
%    \& Sandage, A.  1985, \aj, 90, 473
%\bibitem[Peterson(1976)]{pet76} Peterson, C. J.  1976, \aj, 81, 617
%\bibitem[Rudnick et al.(2003)]{rudnick03} Rudnick, G. et al., 2003, \apj, 599, 847
%\bibitem[Spitzer(1985)]{spi85} Spitzer, L.  1985, Dynamics of
%    Star Clusters, J. Goodman \& P. Hut, Dordrecht: Reidel, 109
%\bibitem[Treu et al.(2003)]{treu03} Treu, T. et al., 2003, \apj, 591, 53

\end{thebibliography}
\end{document}